\documentclass[
twocolumn,
preprintnumbers,
nofootinbib,
showpacs,
prd,aps
]{revtex4-1}

\usepackage{graphicx}
\usepackage{amsmath, amssymb}
\usepackage{bm}
\usepackage{slashed}
\usepackage{color}

\begin{document}

\preprint{KANAZAWA-17-02}

\title{
Longitudinal $W$ boson scattering in a light scalar top scenario
}

\author{Koji Ishiwata and Yuki Yonekura}

\affiliation{Institute for Theoretical Physics, Kanazawa University,
    Kanazawa 920-1192, Japan}

\date{\today}

\begin{abstract}
Scalar tops in the supersymmetric model affect the potential of the
standard model-like Higgs at the quatum level. In light of the
equivalence theorem, the deviation of the potential from the standard
model can be traced by longitudinal gauge bosons.  In this work, high
energy longitudinal $W$ boson scattering is studied in a TeV-scale
scalar top scenario. ${\cal O}$(1--10\%) deviation from the standard
model prediction in the differential cross section is found depending
on whether the observed Higgs mass is explained only by scalar tops or
by additional contributions at a higher scale.
\end{abstract}

\maketitle

\section{Introduction}
\label{sec:intro}

The recent discovery of the Higgs boson confirmed the standard model
(SM) of particle physics~\cite{Aad:2012tfa,Chatrchyan:2012xdj}. Since
then Higgs properties have been measured at the LHC and found to be
consistent with the standard model prediction~\cite{Higg_ATLAS&CMS};
besides, there has been no sign beyond the standard model in the
experiment.  It is widely believed, however, that the standard model
is not the ultimate theory. Superstring theory is one of candidates
for the ``theory of everything''. It requires supersymmetry (SUSY) due
to consistency, which gives rise to lots of phenomenological
consequences beyond the standard model. For example, it provides a
candidate for dark matter, and three gauge coupling constants are
unified at the grand unification scale.  Supersymmetry affects the
Higgs sector too.  In SUSY another Higgs doublet must be introduced
for phenomenologically acceptable Higgs mechanism to work.  In the
supersymmetric Higgs sector, the electroweak symmetry breaking (EWSB)
is induced by renormalization flow of parameters in the Higgs sector,
which is a solution to the origin of the EWSB since in the SM it is
induced by the ad hoc tachyonic Higgs mass term. In spite of such a
drastic extension, the Higgs sector in the supersymmetric model
reduces to the one in the SM below the electroweak scale when
superpartners are much heavier than the electroweak scale. Considering
the current status, {\it i.e.}, no sign of a new particle so far, this
might be the case, and then it might be difficult to observe a clue of
supersymmetry even in future collider experiments.

In such a circumstance, it is worth recalling that the observed
125~GeV Higgs mass cannot be explained in SUSY at tree level. It is
explained by scalar top (``stop'') loop contribution, for example, in
the minimal supersymmetric standard model
(MSSM)~\cite{Haber:1990aw,Ellis:1990nz,Ellis:1991zd,
  Okada:1990vk,Okada:1990gg,Brignole:1992uf}.  This fact indicates
that stop has an impact on the SM Higgs potential at the quantum
level, which is similar to the Higgs sector in classical scale
invariant model. In a simple classical scale invariant model (a) SM
singlet scalar(s) is (are) introduced. They affect the Higgs potential
at quantum level, which induces the EWSB radiatively. In this
framework, the singlet loop determines the curvature of the Higgs
potential around the minimum, {\it i.e.}, the Higgs mass. Although
Higgs properties, such as mass, production and decay rates at collider
experiments, are almost consistent with the SM values, the Higgs
self-couplings are predicted to significantly deviate from the SM
ones~\cite{Dermisek:2013pta,Endo:2015ifa,Hashino:2015nxa}.  This means
that Higgs potential is the same locally around the minimum but not in
a global picture. Such an effect is imprinted in ficticious bosons in
the Higgs doublet, which are absorbed into longitudinal polarization
of the gauge bosons.  While the measurement of the Higgs
self-couplings is one of main goals of the next-generation lepton
collider, {\it e.g.}, the International Linear Collider (ILC), the
deviation from the SM in the Higgs sector can be also probed at the
LHC in the gauge boson scattering process. It is pointed out in
Ref.\,\cite{Endo:2016koi} that the differential cross sections of
longitudinal gauge boson scattering processes $W^+_LW^+_L\to
W^+_LW^+_L$ and $W^+_LW^-_L\to W^+_LW^-_L$ are changed by more than
${\cal O}$(10\%) in the model, which is described by off-shell Higgs.
Namely the discrepancy between the classical scale invariant model and
the SM can be found in off-shell Higgs in the propagator, for which
the longitudinal gauge boson scattering a good probe.

In supersymmetric model, stops are expected to play a role similar to
the singlet scalars. In this paper we analyze the longitudinal gauge
boson scattering in the framework of the supersymmetric
model. Following the analysis in Ref.\,\cite{Endo:2016koi}, we
formulate the leading order amplitudes of the processes and discuss
the deviation from the standard model prediction numerically.

In the study we consider stops with a mass of less than a few
TeV. Such light stop scenario is motivated by naturalness argument,
and part of parameter space of the scenario has already been excluded
by the direct search at the LHC. In Ref.\,\cite{Aad:2015pfx} scalar
top pair production is analyzed in both a simplified model and
phenomenologically tempered SUSY models in conserved R-parity using
Run 1 data. The updated studies at
$\sqrt{s}=13$~TeV~\,\cite{ATLAS:2016jaa,Aaboud:2016tnv,CMS:2016mwj,Khachatryan:2016oia}
have shown that lighter stop mass region $m_{\tilde{t}_1}\lesssim
850~{\rm GeV}$ is excluded at 95\%CL when the lightest neutralino mass
$m_{\tilde{\chi}_1^0}$ is less than about 300~GeV. On the other hand,
$m_{\tilde{t}_1}\gtrsim 400~{\rm GeV}$ and $m_{\tilde{\chi}_1^0}
\gtrsim 300~{\rm GeV}$ (with $m_{\tilde{t}_1}>m_{\tilde{\chi}_1^0}$)
is still allowed. Another possibility is R-parity violation. Without
R-parity the lightest neutralino decays to the standard model
particles, and thus the above analysis cannot be applied. In R-parity
violated scenario, where especially $L_iL_jE^c_k$ or $L_iQ_jD^c_k$
types with light flavor indices exists, the stop mass below 1 TeV is
excluded~\cite{Chatrchyan:2013xsw,Khachatryan:2016ycy}. On the other
hand, in $U^c_iD^c_jD^c_k$ type R-parity violation, stop lighter than
1 TeV has not been excluded~\cite{Aad:2016kww,ATLAS:2016yhq}.  Thus,
various possibilities have yet to be probed for the light stop
scenario.  The naturalness-inspired light stop scenario in the minimal
supersymmetric standard model will be searched at the LHC with more
data (see, {\it e.g.},
Refs.\,\cite{Pierce:2016nwg,Baer:2016bwh,Chala:2017jgg} for recent
studies).  The electroweak precision test and future lepton collider
may be other powerful options for the light stop
search~\cite{Fan:2014axa}. We show that high energy longitudinal gauge
boson scattering is another tool for the indirect search of the
TeV-scale stop. We note that the present work focuses on rather
theoretical study of longitudinal $W$ boson scattering. To discuss the
discovery potential at collider experiments, one needs full simulation
of the process, for example, $pp\to WW jj$, which is not covered in
this paper.  It is known that the observation of high energy (over
TeV) longitudinal gauge boson scattering would be challenging even in
Run 2 at the LHC. We will discuss the issues in the last section,
along with future prospects.

\section{The light scalar top scenario}
\label{sec:scenario}
\setcounter{equation}{0} 

In this paper, we discuss two types of scenarios regarding the origin
of the Higgs mass in the supersymmetric model:
\begin{itemize}
\item[(a)] Higgs mass is explained in the MSSM particle contents
\item[(b)] Other contributions besides the MSSM particles make the
  observed Higgs mass
\end{itemize}
We assume that the other contributions to the Higgs mass are provided
in higher scale than stop mass, {\it e.g.}, heavy vector-like matters
for scinario (b) (see, for example,
Refs.\,\cite{Moroi:1991mg,Moroi:1992zk,Babu:2004xg,Babu:2008ge,Martin:2009bg,Asano:2011zt,Endo:2011mc,Evans:2011uq,Moroi:2011aa,Endo:2011xq,Hisano:2016hni}). To
be concrete, we consider mass spectra $m_{\tilde{t}}\ll \tilde{m}$ for
both cases. Here $m_{\tilde{t}}$ and $\tilde{m}$ are stop mass scale
(defined later) and the mass scale of the rest of superparticles,
respectively.  It is similar to the so-called split supersymmetry
model discussed in Ref.\,\cite{Giudice:2004tc}. In split supersymmetry
gauginos are ${\cal O}(1\mathchar`-10\,{\rm TeV})$, and the other
superparticles are much heavier. In the present discussion we consider
that stops (and the left-handed sbottom) are also around TeV scale.
Just to keep the GUT multiplet structure we assume that the
right-handed stau has TeV mass,\footnote{For example, gauge coupling
  unification is kept at the level of 0.7-1\% for
  $\tilde{m}=10^{6\mathchar`-12}\,{\rm GeV}$ and $m_{\tilde{t}}=1~{\rm
    TeV}$ in one-loop calculation.}  which does not affect the
following analysis. Namely, our discussion comprises the SM-like Higgs
with the scalar top and it is independent of the details of the other
sector.  In Appendix~\ref{app:analyticcheck} we also discuss
$m_{\tilde{t}}\sim \tilde{m}$ case for a reference, which is also
useful for an analytic check of the later calculation. In this paper we
do not argue the naturalness in the Higgs sector but focus on the
consequence of a TeV-scale stop in the gauge boson scattering.

To define the relevant parameters for the Higgs mass, we give the MSSM
superpotential along with soft SUSY breaking terms,
\begin{align}
  &W_{\rm MSSM}=\lambda_t Q_3\!\cdot \!H_u t^c_R +\mu_H H_u\!\cdot\! H_d
  +\cdots \,,
  \\
  &{\cal L}_{\rm soft}=A_t\lambda_t \tilde{Q}_3\!\cdot\! H_u \tilde{t}_R^*
  -m_L^2 |\tilde{Q}_3|^2 -m_R^2|\tilde{t}_R|^2+\cdots\,,
\end{align}
where $Q_3=(t_L,b_L)^T$, $t^c_R$, $H_u=(H^+_u,H^0_u)^T$ and
$H_d=(H^0_d,H^-_d)^T$ are chiral superfields of the third-generation
left-handed quark doublet, right-handed quark singlet (tilded fields
are their superpartners), up-type Higgs doublet, and down-type Higgs
doublet, respectively, and $A\!\cdot\! B \equiv A^T \epsilon B$
($\epsilon=i\sigma_2$). An ellipsis indicates irrelevant terms in our
following discussion. We assume that all parameters are real for
simplicity. In the supersymmetric model, the stop loop contribution has a
significant impact on the SM Higgs mass. In our study, we adopt
renormalization group (RG) method to determine the Higgs
mass~\cite{Okada:1990gg}.  In the reference, the matching conditions at
the scale $\mu \simeq \mu_{\tilde{t}}\sim m_{\tilde{t}}$ are given by
\begin{align}
 &\lambda_{\rm H}^{\rm SM}(\mu_{\tilde{t}})=
  \lambda_{\rm H}^{\rm SM'}(\mu_{\tilde{t}})
  \nonumber \\
  &\qquad + \frac{N_C (y_t^{{\rm SM'}}(\mu_{\tilde{t}}))^4}{(4\pi)^2}
  \Bigl[-\log \Bigl(\frac{\mu^2_{\tilde{t}}}{m_{\tilde{t}}^2}\Bigr)
    +\frac{X_t^2}{m_{\tilde{t}}^2}
    \Bigl(1-\frac{X_t^2}{12m_{\tilde{t}}^2}\Bigr)\Bigr]\,, 
  \label{eq:matching_lambdaH} \\
  &y_t^{\rm SM}(\mu_{\tilde{t}})=y_t^{\rm SM'}(\mu_{\tilde{t}})\,,
  \label{eq:matching_yt}
\end{align}
where $\lambda_{\rm H}^{\rm SM}$ ($y_t^{\rm SM}$) and $\lambda_{\rm
  H}^{\rm SM'}$ ($y_t^{\rm SM'}$) are the Higgs quartic coupling (top
Yukawa coupling) in the energy regions $\mu < m_{\tilde{t}}$ and
$m_{\tilde{t}} \le \mu\, (< \tilde{m})$, respectively. $N_C=3$,
$m_{\tilde{t}}=\sqrt{m_{\tilde{t}_1}m_{\tilde{t}_2}}$
($m_{\tilde{t}_1}$, $m_{\tilde{t}_2}$ are stop masses), and
$X_t=A_t+\mu_H \cot \beta$ ($\tan \beta =\langle H_u^0\rangle/\langle
H_d^0 \rangle$). $\lambda_{\rm H}^{\rm SM}$ must coincide with the
Higgs quartic coupling in the SM. In Eq.\,\eqref{eq:matching_lambdaH}
the second term on the right-hand side is the threshold correction by
integrating out stops.  In the numerical analysis we solve RG
equations for the gauge coupling constants, top Yukawa coupling, and
Higgs quartic coupling.  (In the numerical study we will use more
accurate expression for the condition \eqref{eq:matching_lambdaH}. See
later discussion.) For scenario (a), we need to determine $X_t$ for a
given $m_{L,R}$ to obtain the observed Higgs mass. Thus we solve the
RG equations in the region $m_t \le \mu$ where $m_t$ is the top mass.
We refer to Refs.\,\cite{Buttazzo:2013uya} and \cite{Haber:1993an} for
$m_t \le \mu\le \mu_{\tilde{t}}$ and $\mu_{\tilde{t}} \le \mu \le
\mu_{\rm SUSY}\,(\sim \tilde{m})$, respectively. The RG equations for
$\mu_{\rm SUSY} \le \mu$ are well known, {\it e.g.}, see
Ref.\,\cite{Drees_textbook}. Here matching conditions at $\mu\simeq
\mu_{\rm SUSY}$, $\lambda_{\rm H}^{\rm SM'}(\mu_{\rm SUSY})=
\frac{1}{8}g_Z^2(\mu_{\rm SUSY}) \cos^2 2\beta$, and $y_t^{\rm
  SM'}(\mu_{\rm SUSY})=\lambda_t(\mu_{\rm SUSY})\sin \beta$
($g_Z^2=g'^2+g^2$ where $g'$ and $g$ are the gauge coupling constants
of $U(1)_Y$ and $SU(2)_L$, respectively) should be used. (The
solutions in this region are unnecessary for the computation of the
scattering amplitudes. We use them for a check of the GUT
unification.) We have checked that the obtained Higgs mass is
consistent with the results by using the FeynHiggs
package~\cite{FeynHiggs}; {\it i.e.}, it agrees within about 2 (6)~GeV
in $X_t<0\,(>0)$ region.  This accuracy suffices for leading order
analysis of longitudinal gauge boson scattering discussed below.  On
the other hand, for scenario (b), assuming an additional contribution
to the Higgs quartic coupling at high energy, such as by vector-like
matters, we only need to solve the RG equations in $\mu\le
\mu_{\tilde{t}}$ in the SM particle contents.  In the later analysis,
we will take $\mu_{\tilde{t}}=m_{\tilde{t}}$ and $\mu_{\rm
  SUSY}=\tilde{m}$.

Note that Eq.\,\eqref{eq:matching_lambdaH} corresponds to leading
order computation in the order counting method shown in
Ref.\,\cite{Endo:2016koi}. In the literature an auxiliary expansion
parameter $\xi$ is introduced to define the leading order term for
each physical quantity. Following their analysis, we assign
\begin{align}
  \lambda_{\rm H}^{\rm SM,\,SM'}
  \rightarrow
  \xi^2\, \lambda_{\rm H}^{\rm SM,\,SM'} &\,,
~~~
  y_t^{\rm SM,\,SM'}
  \rightarrow
  \xi^{1/2}\,   y_t^{\rm SM,\,SM'}\, ,
\nonumber \\
g' \rightarrow \xi\,g' &\,,
~~~
g \rightarrow \xi\,g  \,.
\label{eq:xi_assign}
\end{align}
In this assignment any physical quantities, {\it e.g.,} ${\cal P}$,
can be given as ${\cal P}=\xi^n \sum_{i=0}^{\infty}p_i\xi^i$ in
perturbative expansion.  Then we define $p_0$ as the leading
order. Getting back to Eq.\,\eqref{eq:matching_lambdaH}, both first
and second terms in the right-hand side are counted as $\xi^2$, which
means that not only the first term but also the second term is the
leading order. Thus we regard it as the leading order matching
condition. In Eq.\,\eqref{eq:xi_assign}, we have additionally assigned
the $\xi$ counting for $g'$ and $g$ for consistency, which is
discussed later (see Eqs.\,\eqref{eq:Atree} and
\eqref{eq:ASMtree}). With this assignment, we have neglected terms
such as $g^2\lambda_{\rm H}^{\rm SM}$ and $g^4$ in
Eq.\,\eqref{eq:matching_lambdaH}, which are $\xi^4$.  In the following
discussion we use this method to compute the scattering amplitudes at
the leading order.

Before performing the actual calculation, let us estimate the
scattering amplitude.  As pointed out in Ref.\,\cite{Endo:2016koi},
the deviation from the SM in the amplitude high energy gauge boson
scattering is written in terms of the off-shell region of the Higgs
propagator. Although the model is different, scalar tops are expected
to play a role similar to that of the singlet scalars in the
reference.  Then, the deviation from the SM at the leading order
calculation is roughly estimated as
\begin{eqnarray}
  \Delta {\cal A}\sim 
  \frac{N_c y_t^4}{(4\pi)^2}
  \Bigl[\log \Bigl(\frac{p^2}{m_{\tilde{t}}^2}\Bigr)
    +{\cal O}\Bigl(\frac{X_t^2}{p^2}\Bigr)\Bigr]\,,
  \label{eq:DeltaA_est}
\end{eqnarray}
for $|p^2|\gg m_{Z}^2$ ($m_Z$ is the $Z$ boson mass), where $y_t\sim
y_t^{\rm SM}\sim y_t^{\rm SM'}$, and $p$ is the typical momentum of
the scattering process. The logarithmic term, which is from divergent
stop loop diagrams, is the dominant part for $|p^2|\gg
m_{\tilde{t}}^2, X_t^2$, and it can be understood in terms of the RG
flow of the Higgs quartic coupling. However, as emphasized in
Ref.\,\cite{Endo:2016koi}, detailed kinematics, such as energy
dependence or angular distribution, of the scattering process cannot
be described merely in RG computation. In addition, the second term of
the bracket, which cannot be taken into account by RG computation,
may also be comparable to the logarithmic term when $|p^2|\sim
m_{\tilde{t}}^2, X_t^2$. Our main goal is to quantitatively show the
behavior of the gauge boson scattering amplitudes in the existence of
scalar tops in the SUSY model.

\section{Nambu-Goldstone boson scattering }
\label{sec:NGscattering}

\subsection{Equivalence theorem}

\begin{table}[t]
  \begin{center}
    $W^+_LW^+_L$ vs. $G^+G^+$ ($\cos \theta=0$)
  \begin{tabular}{cccccc}
   \hline \hline
   $\sqrt{s}$\,[TeV]  & 0.6 & 1 & 2 & 5 & 10 \\
   \hline
   $[d\sigma/d\cos\theta]_{G^+G^+}$\,[pb]& 9.571 & 3.446 & 0.8614&0.1378&0.03446\\
$[d\sigma/d\cos\theta]_{W^+_LW^+_L}$\,[pb]& 8.361 & 3.286 & 0.8513 & 0.1376 & 0.03444\\
   \hline \hline
   \\
  \end{tabular}
  \end{center}
     \begin{center}
    $W^+_LW^-_L$ vs. $G^+G^-$ ($\cos \theta=0.5$)
  \begin{tabular}{cccccc}
   \hline \hline
   $\sqrt{s}$\,[TeV]  & 0.6 & 1 & 2 & 5 & 10 \\
   \hline
   $[d\sigma/d\cos\theta]_{G^+G^-}$\,[pb]& 1.509 &0.5431& 0.1358& 0.02173& 0.005431 \\
$[d\sigma/d\cos\theta]_{W^+_LW^-_L}$\,[pb]& 1.913& 0.5974& 0.1392& 0.02181& 0.005437 \\
   \hline \hline
  \end{tabular}
    \end{center}
  \caption{\small Differential cross section for $W^+_LW^+_L\to
    W^+_LW^+_L$ and $G^+G^+\to G^+G^+$ (upper) and $W^+_LW^+_L\to
    W^+_LW^+_L$ and $G^+G^-\to G^+G^-$ (lower). }
  \label{table:WWvsGG}
\end{table}

Since we are interested in high energy longitudinal gauge boson
scattering, the equivalence theorem can be applied in our calculation.
The equivalence theorem tells us that the high energy longitudinal
gauge boson ($W^\pm_L$, $Z_L$) corresponds to Nambu-Goldstone (NG)
boson ($G^\pm$, $G^0$). First we will check the validity of the
equivalence theorem quantitatively. To this end we compare the
differential cross section in center-of-mass frame for the processes
$W_L^+W_L^\pm\to W_L^+W_L^\pm$ and $G^+G^\pm\to G^+G^\pm$ in the
SM. The results are summarized in Table \ref{table:WWvsGG}. Here we
use the tree-level analytic formulas given in
Ref.\,\cite{Endo:2016koi} and take the same input parameters, {\it
  i.e.}, $m_W = 80.385$~GeV ($W$ boson mass), $m_Z = 91.1876$~GeV,
$m_h = 125.03$~GeV~\cite{Aad:2014aba,Khachatryan:2014ira}, and $g =
0.65178$. $\theta$ is the scattering angle.  The deviations between
$G^+G^+$ and $W^+_LW^+_L$ ($\cos\theta=0$) are $14$\%, $4.9$\%,
$1.2$\%, $0.19$\% $0.047$\% for center-of-mass energy $\sqrt{s}=0.6$,
1, 2, 5, and 10~TeV, respectively.  On the other hand, for $W^+W^-$
($G^+G^-$) scattering, the deviations are 21\%, 10\%, 2.5\%, 0.40\%,
and 0.10\% in the same $\sqrt{s}$ but for $\cos \theta = 0.5$. It is
seen that the deviation gets smaller for larger $\sqrt{s}$ as
expected.  In the backward region, on the other hand, the differential
cross section is suppressed due to a cancellation in the tree-level
amplitude. In such a region the other one-loop contributions besides
(scalar) top and bottom, {\it i.e.}, electroweak corrections,
including the Sudakov logarithm~\cite{Fadin:1999bq,Kuhn:2011mh},
become numerically important~\cite{Denner:1997kq}. It is discussed in
Ref.\,\cite{Denner:1997kq} that the finite decay width of $W$ bosons
must be taken into account by using the complex mass
scheme~\cite{Denner:2006ic} or considering the actual decay chains of
$W$ bosons~\cite{Accomando:2006hq} for consistent calculation. Since
those issues are beyond the scope of the present study, we discard
backward region.\footnote{Here note that we do not insist that the
  forward region is effective for our study. As we will see later, it
  is dominated by $\gamma$ and $Z$ boson exchange diagrams and not so
  efficient for seeing the deviation from the SM. (Central or
  semicentral regions are more promising.)}

In the later numerical analysis, we discuss the differential cross
section in the SM and the supersymmetric model at the level of ${\cal
  O}$(1--10\%).  Thus, to substitute the NG boson scattering for
longitudinal $W$ boson scattering at less than about 0.1\% we will
mainly consider $\sqrt{s}\ge 2~{\rm TeV}$. Note that the number of
events where the $W$ boson system has the invariant mass over 2 TeV is
expected to be limited even in Run 2 at the LHC. As mentioned in the
Introduction, we try to show a potential of $WW$ scattering for the
study of beyond the SM in a long-term period, considering in the
future a high energy frontier experiment, such as the Future Circular
Collider.

\subsection{Scattering amplitudes}

In this subsection we will calculate the $G^+G^\pm\to G^+G^\pm$
scattering amplitude.  The interaction terms which are relevant for
the scattering processes in our current setup are
\begin{align}
  &{\cal L} = -  \lambda_{\rm H}^{\rm SM'}|G^+ G^-|^2 
  - y_t^{\rm SM'} X_t (
  \tilde{b}^*_L\tilde{t}_RG^-+\tilde{b}_L\tilde{t}^*_RG^+) \nonumber \\ 
 & -\Bigl[(y_t^{\rm SM'})^2
    -\frac{1}{2}g_Z^2\Bigl(-\frac{1}{2}+\frac{2}{3}\sin^2\theta_W\Bigr)
    \cos2\beta \Bigr]|\tilde{b}_L|^2 |G^+|^2
  \nonumber \\
   & -\Bigl[(y_t^{\rm SM'})^2
    +\frac{1}{2}g_Z^2\,\frac{2}{3}\sin^2\theta_W
    \cos2\beta \Bigr]|\tilde{t}_R|^2 |G^+|^2\,,
  \label{eq:L}
\end{align}
where the couplings are defined in Eqs.\,\eqref{eq:matching_lambdaH}
and \eqref{eq:matching_yt} and $\theta_W$ is the Weinberg angle.  In
the following calculation, we take $\overline{\rm MS}$ scheme in
dimensional regularization and use LoopTools~\cite{Hahn:1998yk} for
the numerical study.

\begin{figure}[t]
  \begin{center}
    \includegraphics[scale=0.5]{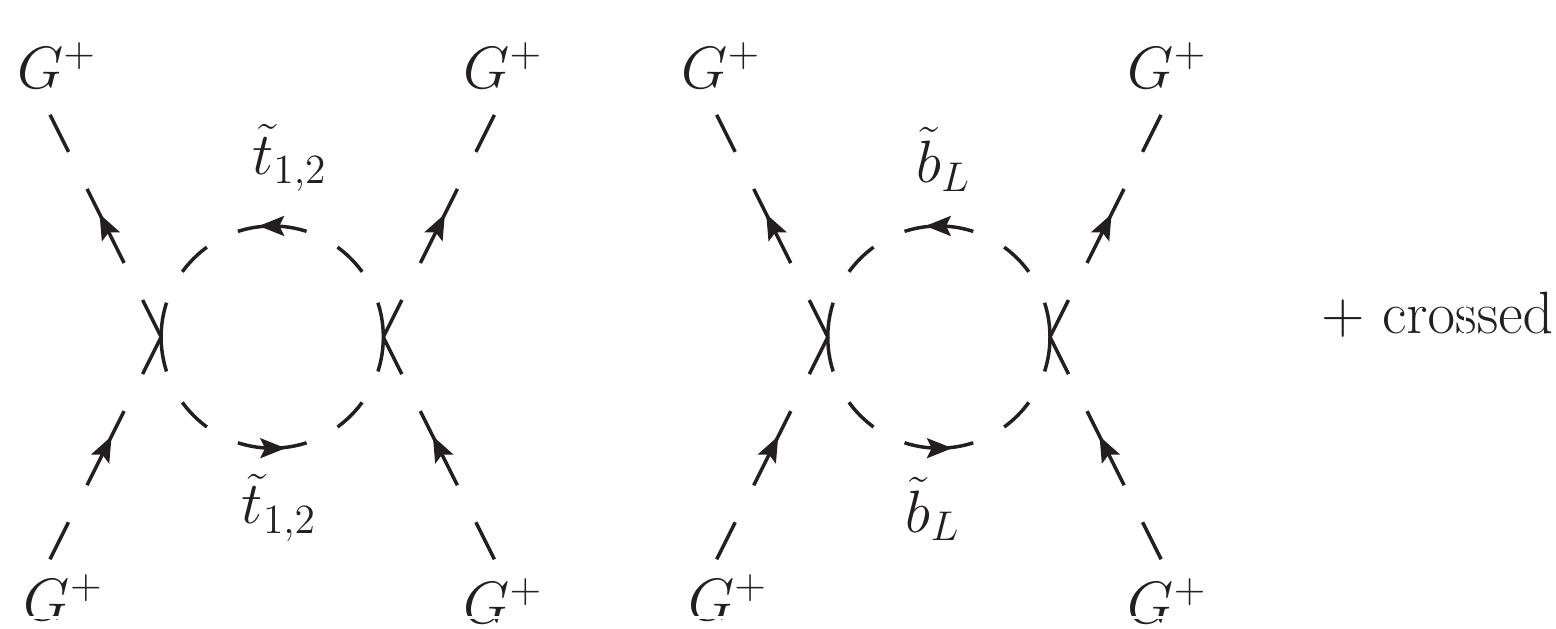}
    \includegraphics[scale=0.5]{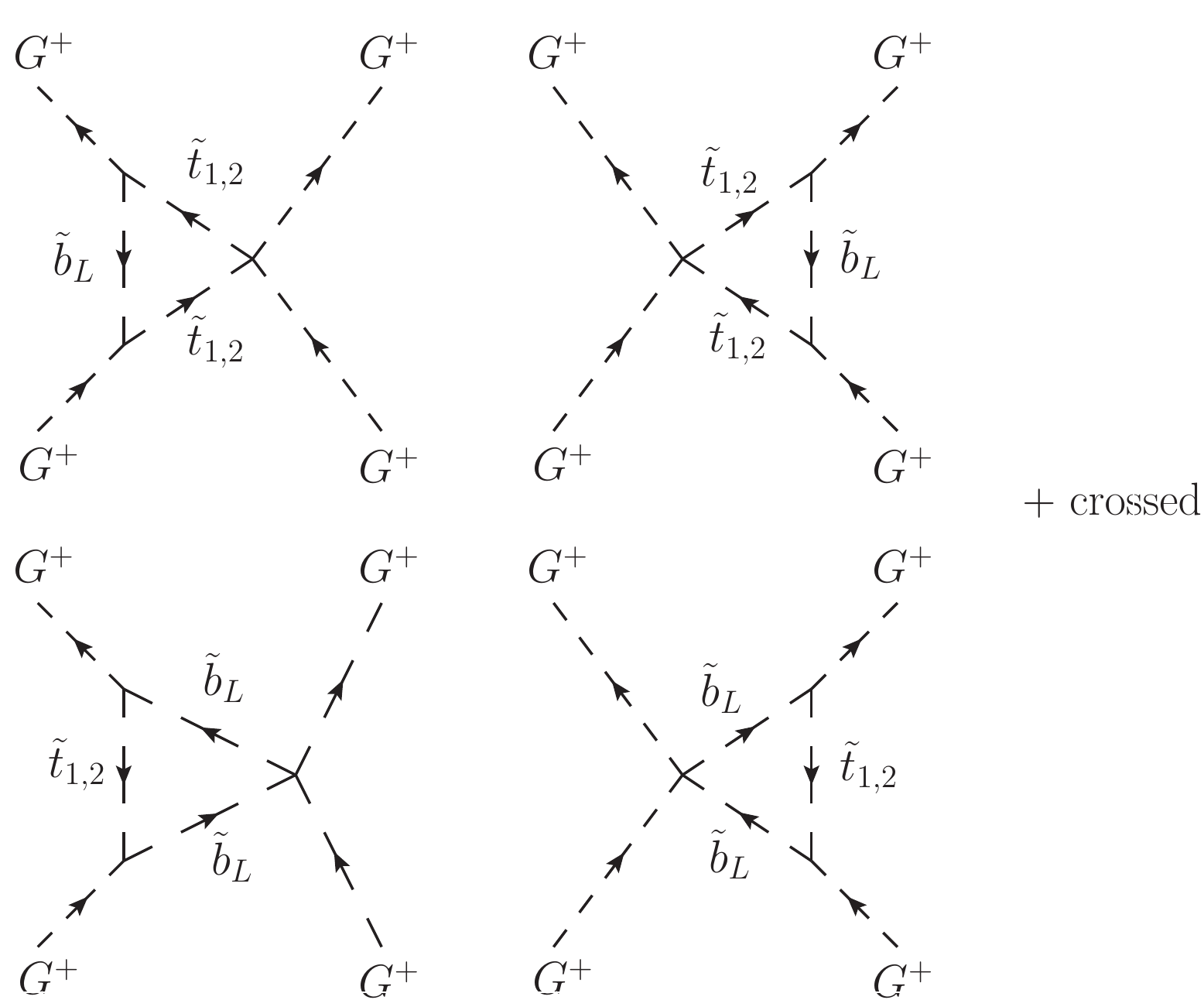}
    \includegraphics[scale=0.5]{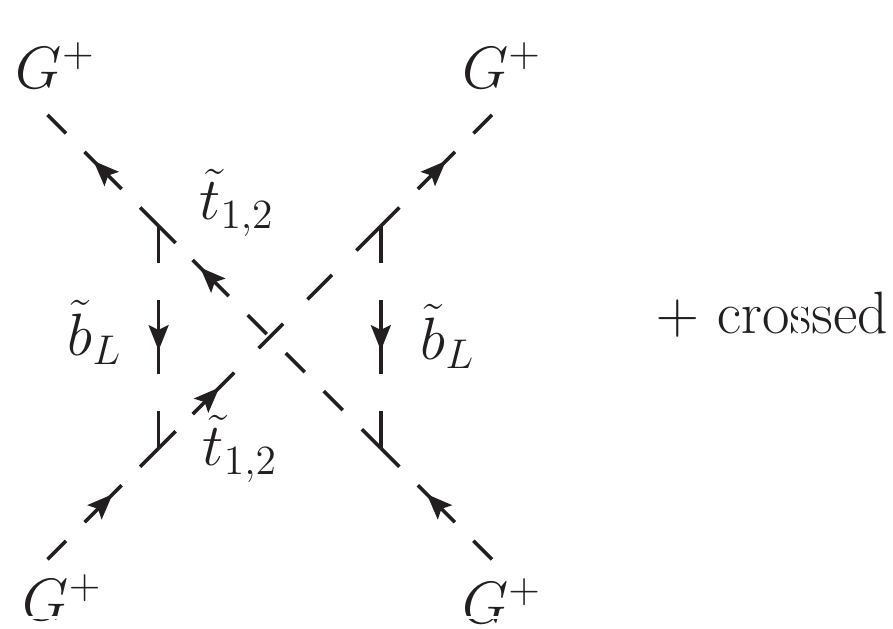}
   \end{center}
   \caption{Stop-sbottom loop diagrams which induce $G^+G^+$
     scattering. Time flows upward, and ``crossed'' means crossed
     diagram for final-state bosons. Circle-type, triangle-type, and
     crossed box-type diagrams correspond to ${\cal
       A}_{G^+G^+}^{\tilde{t}\mathchar`-{\tilde{b}},{\rm cir}}$,
     ${\cal A}_{G^+G^+}^{\tilde{t}\mathchar`-{\tilde{b}},{\rm tri}}$,
     and ${\cal A}_{G^+G^+}^{\tilde{t}\mathchar`-{\tilde{b}},{\rm
         box}}$ in Eq.\,\eqref{eq:stoploop}, respectively. }
   \label{fig:stoploop}
\end{figure}

Let us discuss $G^+G^+\to G^+G^+$ scattering first.  The scattering
amplitudes in the supersymmetric model and the SM are given by the
form
\begin{align}
  &{\cal A}_{G^+G^+}=
  {\cal A}_{G^+G^+}^{\rm tree} + {\cal A}_{G^+G^+}^{t\mathchar`-b}+
  {\cal A}_{G^+G^+}^{\tilde{t}\mathchar`-{\tilde{b}}}\,,
\label{eq:AG+G+}\\
  &{\cal A}_{G^+G^+}^{\rm SM}=
    {\cal A}_{G^+G^+}^{\rm SM,\,tree} +
    {\cal A}_{G^+G^+}^{{\rm SM},\,t\mathchar`-b}\,,
\end{align}
where ``tree'', ``$t\mathchar`-b$'', and
``$\tilde{t}\mathchar`-{\tilde{b}}$'' indicate the tree-level
amplitude, top-bottom loop amplitude, and stop-sbottom loop amplitude,
respectively, which are given by
\begin{align}
  {\cal A}_{G^+G^+}^{\rm tree}
  &= - 4\lambda_{\rm H}^{\rm SM'}
  -\frac{g_Z^2}{2}\Bigl[\frac{t}{u}+\frac{u}{t}+1\Bigr]\,,
  \label{eq:Atree}\\
  {\cal A}_{G^+G^+}^{\rm SM,\,tree}
  &= - 4\lambda_{\rm H}^{\rm SM}
  -\frac{g_Z^2}{2}\Bigl[\frac{t}{u}+\frac{u}{t}+1\Bigr]\,,
  \label{eq:ASMtree} 
\end{align}
\begin{align}
  {\cal A}_{G^+G^+}^{t\mathchar`-b}
  &= -\frac{2N_c (y_t^{\rm SM'})^4}{(4\pi)^2}
  \Bigl[B_0(t,m_t^2,m_t^2)+B_0(u,m_t^2,m_t^2)\Bigr]\,,\\
  \label{eq:A^top}
  {\cal A}_{G^+G^+}^{{\rm SM},\,t\mathchar`-b}
  &= -\frac{2N_c (y_t^{\rm SM})^4}{(4\pi)^2}
  \Bigl[B_0(t,m_t^2,m_t^2)+B_0(u,m_t^2,m_t^2)\Bigr]\,,
\end{align}
where $t=(p_1-k_1)^2$, $u=(p_1-k_2)^2$ ($p_i$ and $k_i$ ($i,j=1,2$)
are momenta of incident and scattered particles, respectively), and
$B_0$ is the loop function defined in Eq.\,(B.5) in
Ref.\,\cite{Endo:2016koi} without $1/\bar{\epsilon}$.  The couplings
are renormalized ones and their $\mu$ dependence is implicit.  Here we
have taken the leading terms in the $|t|,|u|\gg m_Z^2$ limit.  ${\cal
  A}_{G^+G^+}^{\tilde{t}\mathchar`-{\tilde{b}}}$ consists of three
types of diagrams, circle, triangle and box types, which are shown in
Fig.\,\ref{fig:stoploop}. We can derive them straightforwardly as
\begin{eqnarray}
  {\cal A}_{G^+G^+}^{\tilde{t}\mathchar`-{\tilde{b}}} = 
  {\cal A}_{G^+G^+}^{\tilde{t}\mathchar`-{\tilde{b}},{\rm cir}}+
  {\cal A}_{G^+G^+}^{\tilde{t}\mathchar`-{\tilde{b}},{\rm tri}}+
  {\cal A}_{G^+G^+}^{\tilde{t}\mathchar`-{\tilde{b}},{\rm box}}\,,
  \label{eq:stoploop}
\end{eqnarray}
with 
\begin{widetext} 
\begin{align}
  {\cal A}_{G^+G^+}^{\tilde{t}\mathchar`-{\tilde{b}},{\rm cir}}
  =& \frac{N_C (y_t^{\rm SM'})^4}{(4\pi)^2}
  \Bigl[
    s^4_{\theta_t}B_0(t,m_{\tilde{t}_1}^2,m_{\tilde{t}_1}^2)+
    c^4_{\theta_t}B_0(t,m_{\tilde{t}_2}^2,m_{\tilde{t}_2}^2)+
    2s^2_{\theta_t}c^2_{\theta_t}B_0(t,m_{\tilde{t}_1}^2,m_{\tilde{t}_2}^2)
    + B_0(t,m_{\tilde{b}_L}^2,m_{\tilde{b}_L}^2) \Bigr]
  \nonumber \\ 
      &\qquad \qquad \qquad
  +(t\to u ~{\rm term})\,,
  \label{eq:cir}
  \\
    {\cal A}_{G^+G^+}^{\tilde{t}\mathchar`-{\tilde{b}},{\rm tri}}
    =& \frac{2N_C (y_t^{\rm SM'})^4X_t^2}{(4\pi)^2}
 \Bigl[
   s^4_{\theta_t}C_0(0,t,0,m_{\tilde{b}_L}^2,m_{\tilde{t}_1}^2,m_{\tilde{t}_1}^2)+
   c^4_{\theta_t}C_0(0,t,0,m_{\tilde{b}_L}^2,m_{\tilde{t}_2}^2,m_{\tilde{t}_2}^2)+
   2s^2_{\theta_t}c^2_{\theta_t}
   C_0(0,t,0,m_{\tilde{b}_L}^2,m_{\tilde{t}_1}^2,m_{\tilde{t}_2}^2)
   \nonumber \\ 
   &\qquad \qquad \qquad
   +s^2_{\theta_t}C_0(0,t,0,m_{\tilde{t}_1}^2,m_{\tilde{b}_L}^2,m_{\tilde{b}_L}^2)+
  c^2_{\theta_t}C_0(0,t,0,m_{\tilde{t}_2}^2,m_{\tilde{b}_L}^2,m_{\tilde{b}_L}^2) \Bigr]
 +(t\to u ~{\rm term}) \,,
 \label{eq:tri}
 \\
   {\cal A}_{G^+G^+}^{\tilde{t}\mathchar`-{\tilde{b}},{\rm box}}
   =&\frac{2N_C (y_t^{\rm SM'})^4X_t^4}{(4\pi)^2}
   \Bigl[
     s^4_{\theta_t}
     D_0(0,0,0,0,u,t,
     m_{\tilde{b}_L}^2,m_{\tilde{t}_1}^2,m_{\tilde{b}_L}^2,m_{\tilde{t}_1}^2)+
     c^4_{\theta_t}
     D_0(0,0,0,0,u,t,
     m_{\tilde{b}_L}^2,m_{\tilde{t}_2}^2,m_{\tilde{b}_L}^2,m_{\tilde{t}_2}^2)
     \nonumber \\
     &\qquad \qquad \qquad +2s^2_{\theta_t}c^2_{\theta_t}
     D_0(0,0,0,0,u,t,
     m_{\tilde{b}_L}^2,m_{\tilde{t}_1}^2,m_{\tilde{b}_L}^2,m_{\tilde{t}_2}^2)
     \Bigr]\,.
   \label{eq:box}
\end{align}
\end{widetext}
Loop functions $C_0$ and $D_0$ are those defined in
Ref.\,\cite{Hahn:1998yk}. $m_{\tilde{b}_L}$ is the left-handed sbottom
mass. Since we consider that the right-handed sbottom mass is much
larger than the third-generation left-handed squark mass, the lighter
sbottom is mostly composed of $b_L$; thus, $m_{\tilde{b}_L}\simeq
m_L$. $\theta_t$ is the mixing angle in stop sector defined as
$(\tilde{t}_1,\tilde{t}_2)^T=Z\, (\tilde{t}_L,\tilde{t}_R)^T$ with
orthogonal matrix $Z_{11}=\cos \theta_t \equiv c_{\theta_t}$,
$Z_{12}=\sin \theta_t \equiv s_{\theta_t}$. To be consistent with
$\xi$ expansion analysis, we have omitted terms such as $(y_t^{\rm
  SM'})^2g_Z^2$ and $g_Z^4$ in Eq.\,\eqref{eq:cir}, which are $\xi^3$
and $\xi^4$, respectively, in $\xi$ expansion.  We note that the
explicit $\mu$ dependence coming from $B_0$ function is canceled at
the leading order by the RG flow of $\lambda_{\rm H}^{\rm SM}$
($\lambda_{\rm H}^{\rm SM'}$).  Since our goal is to compute the
deviation at leading order in $\xi$ expansion, we take
$\mu=m_{\tilde{t}}$ in the amplitude hereafter.

Another scattering amplitude for the process $G^+G^-\to G^+ G^-$ can
be obtained by replacing the Mandelstam variable $u$ by $s$.

Before going to the numerical analysis, let us check low- and
high-energy limits. In the low-energy limit, the amplitudes ${\cal
  A}_{G^+G^+}$ and ${\cal A}_{G^+G^-}$ should coincide with those in
the SM. To see this we define $\Delta {\cal A}_{G^+G^\pm}$
\begin{align}
 \Delta {\cal A}_{G^+G^\pm}= {\cal A}_{G^+G^\pm}-{\cal A}_{G^+G^\pm}^{\rm SM}\,.
\end{align}
Then, using the matching conditions \eqref{eq:matching_lambdaH} and
\eqref{eq:matching_yt}, they are simply given by
\begin{align}
  \Delta {\cal A}_{G^+G^\pm}=
  {\cal A}_{G^+G^\pm}^{\tilde{t}\mathchar`-{\tilde{b}}}
  +\frac{4N_C (y_t^{\rm SM}(m_{\tilde{t}}))^4}{(4\pi)^2}
  \frac{X_t^2}{m_{\tilde{t}}^2}
    \Bigl(1-\frac{X_t^2}{12m_{\tilde{t}}^2}\Bigr)\,,
    \label{eq:DeltaA}
\end{align}
In the low-energy limit, $s,|t|,|u|\ll m_{\tilde{t}}^2$ (but
$s,|t|,|u|\gg m_Z^2$), and taking $m_{\tilde{t}_1}\simeq
m_{\tilde{t}_2}\simeq m_L \simeq m_{\tilde{t}}$\, the stop-sbottom loop
contribution behaves as
\begin{align}
  &{\cal A}_{G^+G^\pm}^{\tilde{t}\mathchar`-{\tilde{b}},{\rm cir}}
  \longrightarrow
  \frac{4N_C (y_t^{\rm SM'}(\mu))^4}{(4\pi)^2}
  \log \Bigl(\frac{\mu^2}{m_{\tilde{t}}^2}\Bigr)\Bigr|_{\mu=m_{\tilde{t}}}\,,
  \\
  &{\cal A}_{G^+G^\pm}^{\tilde{t}\mathchar`-{\tilde{b}},{\rm tri}}
  \longrightarrow
  \frac{8N_C (y_t^{\rm SM'}(\mu))^4X_t^2}{(4\pi)^2}
  \Bigl(-\frac{1}{2m^2_{\tilde{t}}}\Bigr)\Bigr|_{\mu=m_{\tilde{t}}}\,,
  \\
  &{\cal A}_{G^+G^\pm}^{\tilde{t}\mathchar`-{\tilde{b}},{\rm box}}
  \longrightarrow
   \frac{2N_C (y_t^{\rm SM'}(\mu))^4X_t^4}{(4\pi)^2}
  \Bigl(\frac{1}{6m^4_{\tilde{t}}}\Bigr)\Bigr|_{\mu=m_{\tilde{t}}}\,,
\end{align}
which leads to
\begin{eqnarray}
{\cal A}_{G^+G^\pm}^{\tilde{t}\mathchar`-{\tilde{b}}}
\longrightarrow
-  \frac{4N_C (y_t^{\rm SM}(m_{\tilde{t}}))^4}{(4\pi)^2}
  \Bigl[
    \frac{X_t^2}{m^2_{\tilde{t}}}\Bigl(1-\frac{X_t^2}{12m^2_{\tilde{t}}}\Bigr)
    \Bigr] \,.
  \label{eq:A^stop_lowenergylim}
\end{eqnarray}
Thus $\Delta {\cal A}_{G^+G^-}\to 0$, which means that the amplitude
asymptotically approaches  the SM one in the low-energy limit as
expected.

In numerical calculation $m_{\tilde{t}_1}\simeq m_{\tilde{t}_2}\simeq
m_{\tilde{t}}$ is not always satisfied. Therefore, in the later
analysis, we use the following expressions instead of
Eqs.\,\eqref{eq:DeltaA} and \eqref{eq:matching_lambdaH};
\begin{align}
  &\Delta {\cal A}_{G^+G^\pm}=
{\cal A}_{G^+G^\pm}^{\tilde{t}\mathchar`-{\tilde{b}}}-
{\cal A}_{G^+G^\pm}^{\tilde{t}\mathchar`-{\tilde{b}}}|_{s\to 0} \,,     
\label{eq:DeltaAnum}\\
 &\lambda_{\rm H}^{\rm SM}(\mu_{\tilde{t}})=
\lambda_{\rm H}^{\rm SM'}(\mu_{\tilde{t}})
-\frac{1}{4}{\cal A}_{G^+G^\pm}^{\tilde{t}\mathchar`-{\tilde{b}}}|_{s\to 0} \,.      
\end{align}

In the high-energy limit $s,|t|,|u|\gg m_{\tilde{t}},|X_t|$, on the other
hand,
\begin{widetext}
\begin{align}
  {\cal A}_{G^+G^+}^{\tilde{t}\mathchar`-{\tilde{b}}}
  \longrightarrow 
  &\frac{4N_C (y_t^{\rm SM'}(m_{\tilde{t}}))^4}{(4\pi)^2}
  \Bigl[\log \Bigl(\frac{m^2_{\tilde{t}}}{\sqrt{tu}}\Bigr)-2
    +\frac{m^2_{\tilde{t}}}{t}\Bigl(\log\frac{m^2_{\tilde{t}}}{-t}-1\Bigr)
    +\frac{m^2_{\tilde{t}}}{u}\Bigl(\log\frac{m^2_{\tilde{t}}}{-u}-1\Bigr)
    \nonumber \\
    &\qquad \qquad \qquad 
    +\frac{X_t^2}{2t} \log^2\Bigl(\frac{m^2_{\tilde{t}}}{-t}\Bigr)
    +\frac{X_t^2}{2u} \log^2\Bigl(\frac{m^2_{\tilde{t}}}{-u}\Bigl)
    \Bigr]
  +{\cal O}\Bigl(
  \frac{m^4_{\tilde{t}},X_t^4}{|t|^2,|u|^2}\Bigr)
  \,,
  \label{eq:stoploop_highenergylim_GpGp}
  \\
  {\cal A}_{G^+G^-}^{\tilde{t}\mathchar`-{\tilde{b}}}
  \longrightarrow 
  &\frac{2N_C (y_t^{\rm SM'}(m_{\tilde{t}}))^4}{(4\pi)^2}
  \Bigl[\log \Bigl(\frac{m^2_{\tilde{t}}}{\sqrt{-st}}\Bigr)-2
    +\frac{m^2_{\tilde{t}}}{s}\Bigl(\log\frac{m^2_{\tilde{t}}}{s}-1\Bigr)
    +\frac{m^2_{\tilde{t}}}{t}\Bigl(\log\frac{m^2_{\tilde{t}}}{-t}-1\Bigr)
    -\frac{i\pi}{2}\Bigl(1-2\frac{m^2_{\tilde{t}}}{s}\Bigr)
    \nonumber \\
    &\qquad \qquad \qquad
    +\frac{X_t^2}{2s} \Bigl\{
      \log^2\Bigl(\frac{m^2_{\tilde{t}}}{s}\Bigl)
      -\pi^2+2i\pi\log\Bigl(\frac{m^2_{\tilde{t}}}{s}\Bigl)
      \Bigr\}
    +\frac{X_t^2}{2t} \log^2\Bigl(\frac{m^2_{\tilde{t}}}{-t}\Bigr)
    \Bigr]
  +{\cal O}\Bigl(
  \frac{m^4_{\tilde{t}},X_t^4}{s^2,|t|^2}\Bigr)
  \,.
  \label{eq:stoploop_highenergylim_GpGm}
\end{align}
\end{widetext}
The first line on the right-hand side comes from circle diagram, which
agrees with the native estimation \eqref{eq:DeltaA_est} and can be
understood in terms of the RG flow of the Higgs quartic coupling.
Meanwhile, the others are derived in the explicit calculation of
Feynman diagrams, which cannot be described by the RG equations and
are necessary ingredients for the numerical analysis of the scattering
processes.

\section{Numerical results}
\label{sec:results}

\begin{figure*}[t]
  \begin{center}
    \includegraphics[scale=0.8]{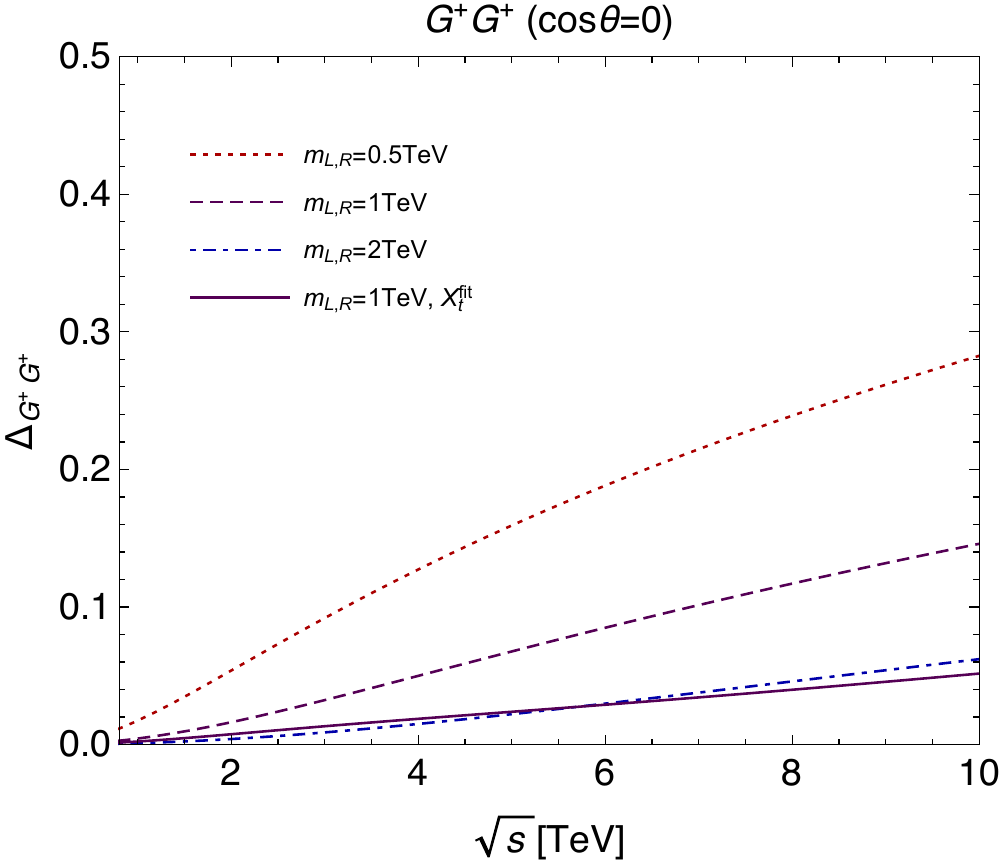}
    \includegraphics[scale=0.8]{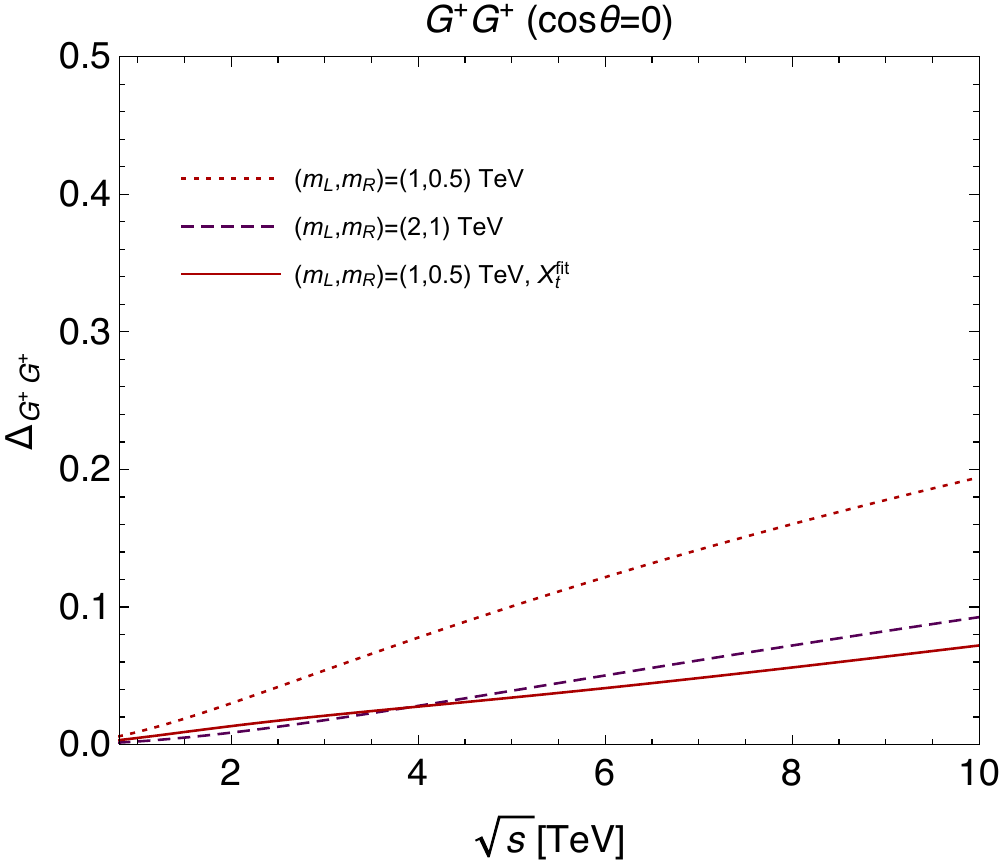}
    \end{center}
  \caption{Deviation from the SM for the $G^+G^+$ scattering process,
    defined in Eq.\,\eqref{eq:Delta_GG}, as a function of
    center-of-mass energy $\sqrt{s}$.  The scattering angle is taken
    as $\cos \theta=0$. (Left) $m_L=m_R=0.5$ (red dotted), 1 (purple
    dashed), and 2 (blue dot-dashed) TeV with $X_t=0.5m_L$ at
    $\mu=m_{\tilde{t}}$. In $X_t^{\rm fit}$ (purple solid) line,
    $X_t=1.82$ TeV, which gives the observed Higgs mass for
    $m_L=m_R=1$~TeV, is taken.  (Right) $m_L=2m_R=1$ (red dotted) and
    2 (purple dashed) TeV with $X_t=0.5m_L$ at $\mu=m_{\tilde{t}}$. In
    $X_t^{\rm fit}$ (red solid) line, $X_t=1.45$ TeV, which gives the
    observed Higgs mass, is taken.}
   \label{fig:Delta-rs_GpGp}
\end{figure*}

\begin{figure*}[t]
  \begin{center}
    \includegraphics[scale=0.8]{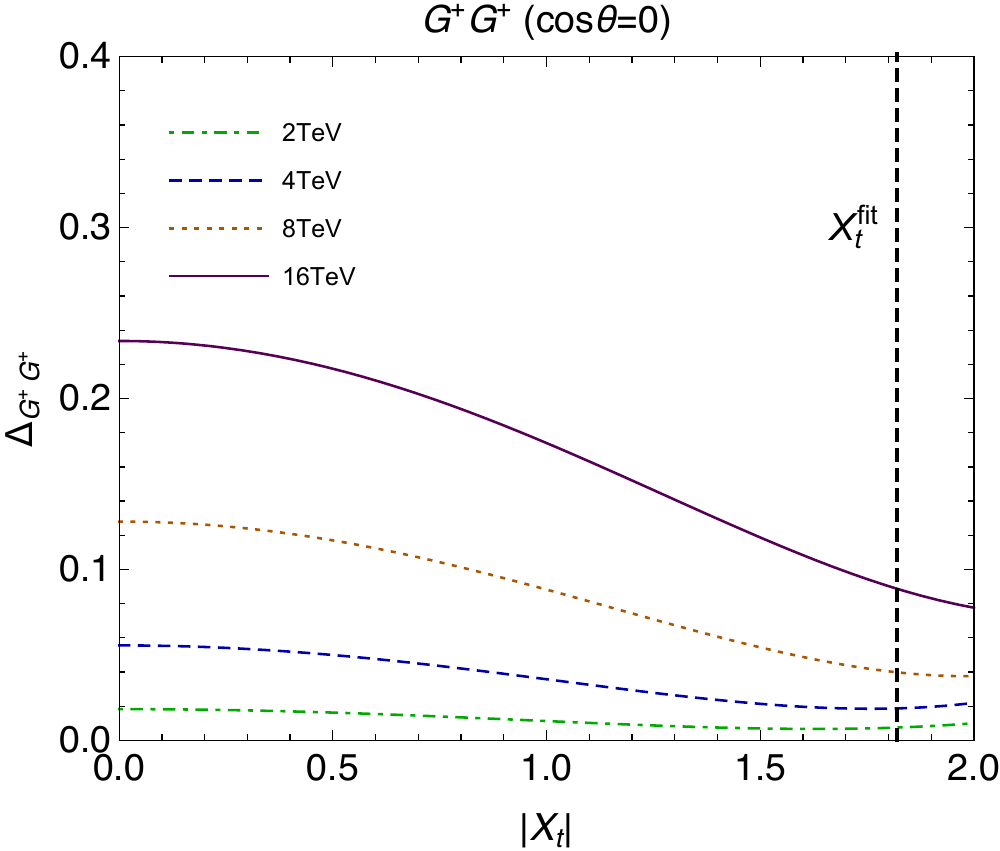}
    \includegraphics[scale=0.8]{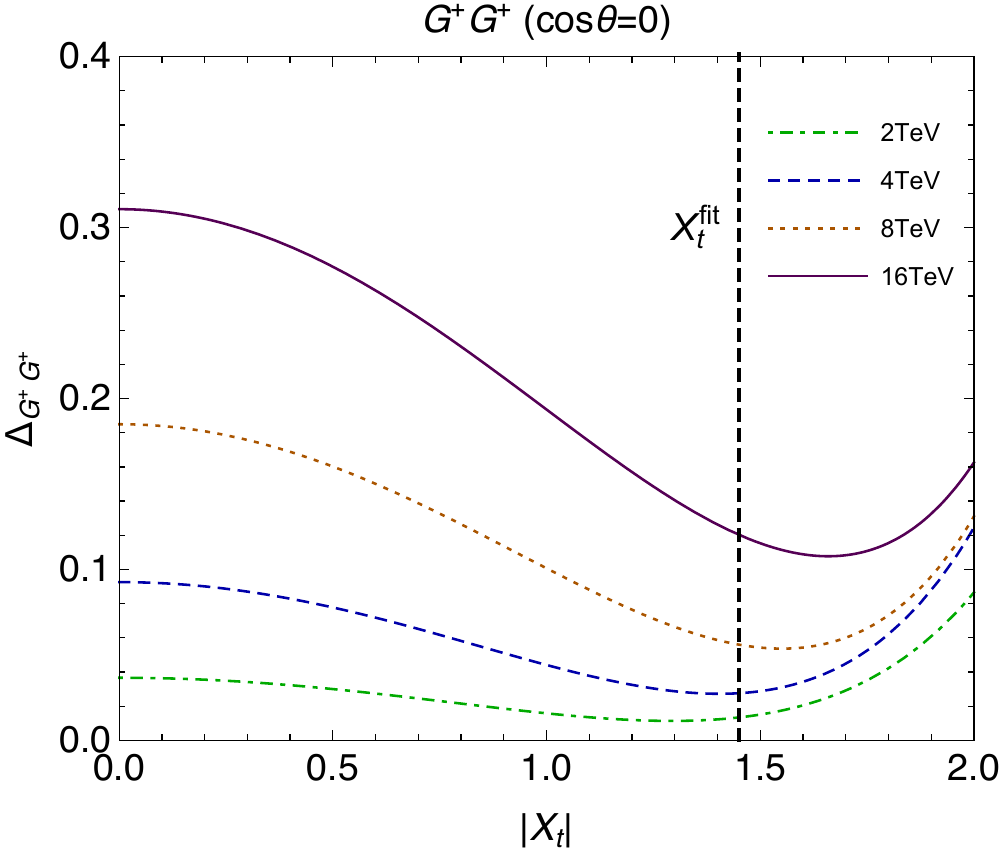}
    \end{center}
  \caption{$X_t$ and $\sqrt{s}$ dependence of the amplitude for
    $G^+G^+$ process. $m_L=m_R=1$~TeV (left) and $m_L=2m_R=1$~TeV
    (right) are taken. Line contents are $\sqrt{s}=2$ (green
    dot-dashed), $4$ (blue dashed), $8$ (orange dotted), and
    $16$~TeV (purple solid). Locations of $X_t^{\rm fit}$ are also
    indicated. }
   \label{fig:Delta-Xt_GpGp}
\end{figure*}

Now we are ready to give the numerical result.  To this end, we use
the quantity:
\begin{align}
  \Delta_{G^+G^\pm}
  = \frac{|{\cal A}_{G^+G^\pm}|^2-|{\cal A}_{G^+G^\pm}^{\rm SM}|^2}
  {|{\cal A}_{G^+G^\pm}^{\rm SM}|^2}\,,
  \label{eq:Delta_GG}
\end{align}
which corresponds to the deviation from the SM for the differential
cross section.

\begin{figure*}[t]
  \begin{center}
    \includegraphics[scale=0.8]{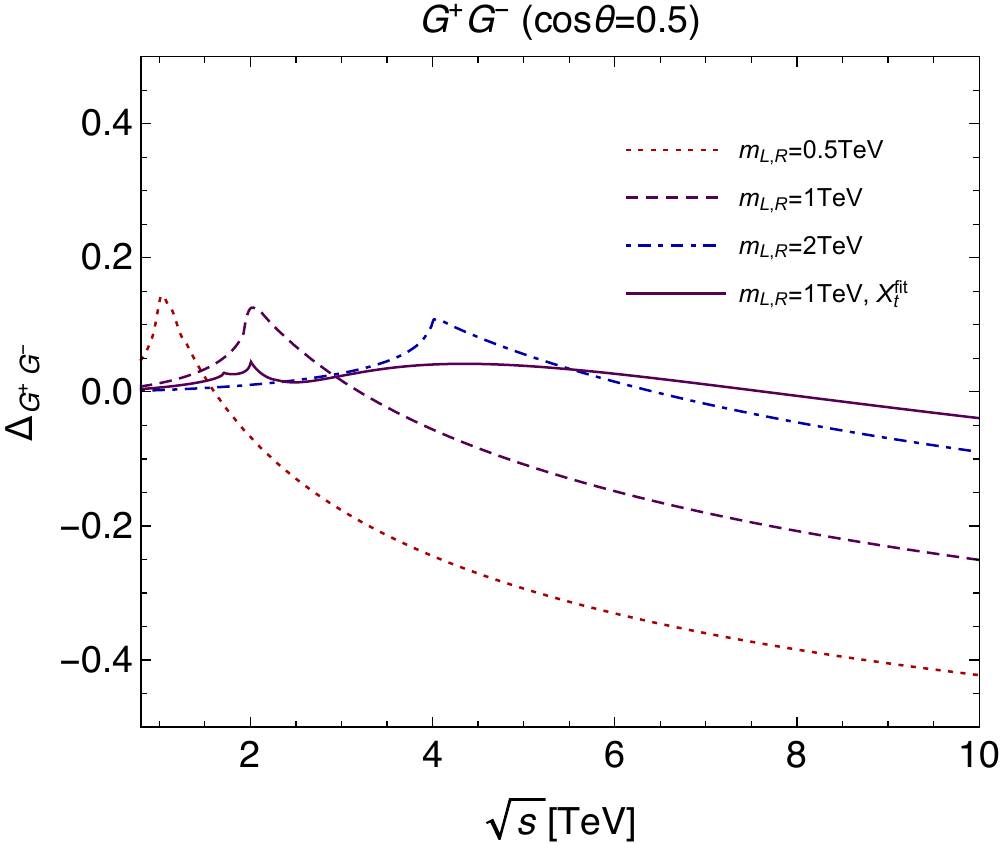}
    \includegraphics[scale=0.8]{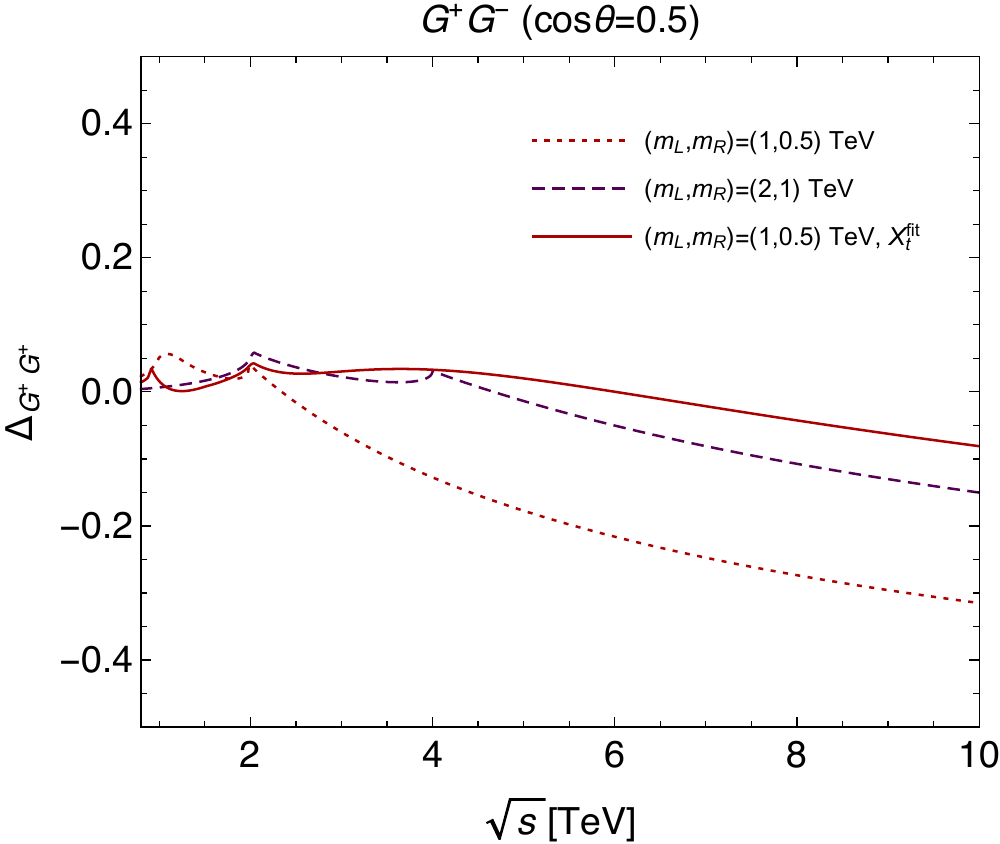}
    \end{center}
  \caption{Same as Fig.\,\ref{fig:Delta-rs_GpGp} but for $G^+G^-$
    scattering process taking $\cos\theta=0.5$. }
   \label{fig:Delta-rs_GpGm}
\end{figure*}

\begin{figure*}[t]
  \begin{center}
    \includegraphics[scale=0.8]{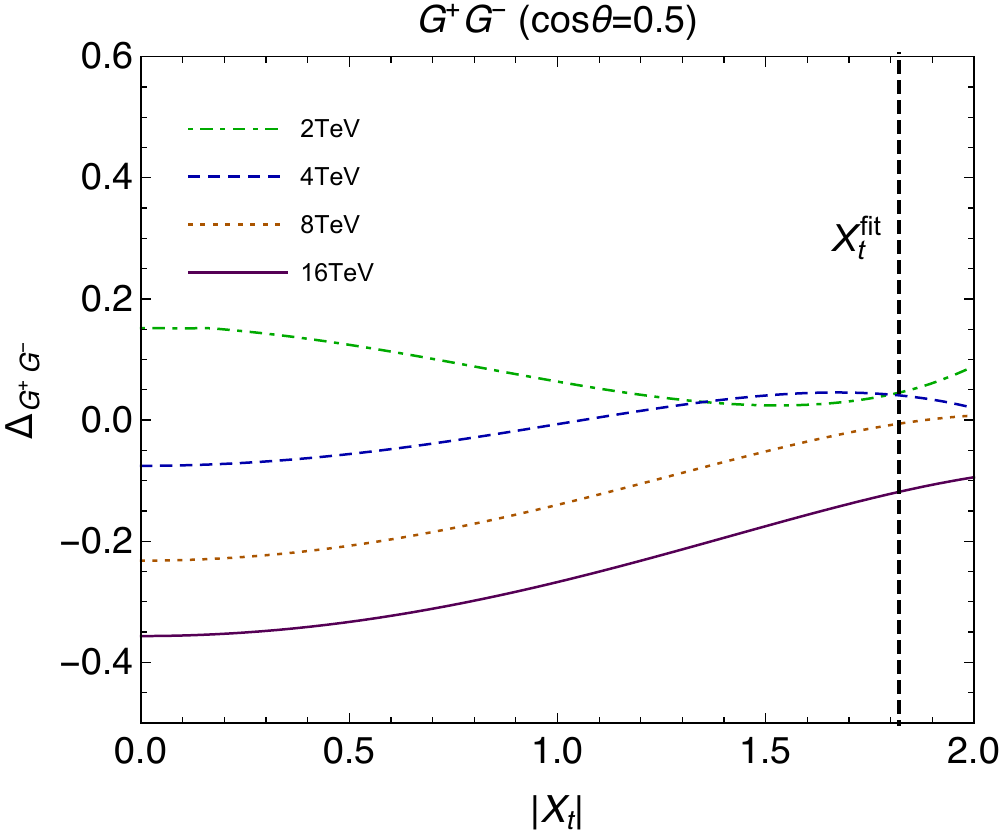}
    \includegraphics[scale=0.8]{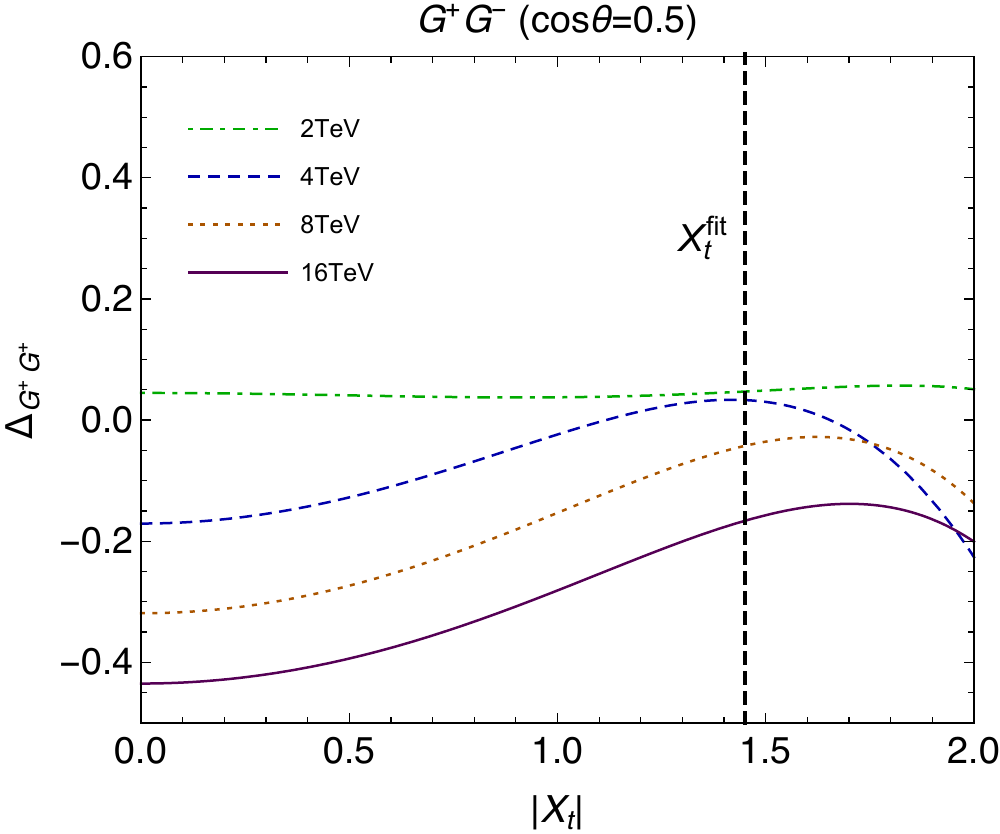}
    \end{center}
  \caption{Same as Fig.\,\ref{fig:Delta-Xt_GpGp} but for $G^+G^-$
    process taking $\cos\theta=0.5$. }
   \label{fig:Delta-Xt_GpGm}
\end{figure*}

Fig.\,\ref{fig:Delta-rs_GpGp} shows the result for the $G^+G^+$
process as a function of $\sqrt{s}$ for the fixed scattering angle,
$\cos \theta=0$. We take $m_L=m_R=0.5$, 1, and 2 TeV (left) and
$m_L=2m_R=1$ and 2 TeV (right) with $X_t=0.5m_L$ at
$\mu=m_{\tilde{t}}$, which corresponds to scenario (b) discussed in
Sec.\,\ref{sec:scenario}.  Roughly speaking, left (right) panel covers
the situation of the degenerate (split) mass spectrum in the stop
sector.  For scenario (a), the results for $m_L=m_R=1$ TeV with
$X_t=X_t^{\rm fit}=1.82~{\rm TeV}$ (left), $m_L=2m_R=1$ TeV with
$X_t=X_t^{\rm fit}=1.45~{\rm TeV}$ (right) are given. Here we omit
another larger value of $|X_t|$ to give the observed Higgs mass since
it would not be phenomenologically acceptable due to vacuum
instability bound $X_t
/\sqrt{m_{\tilde{t}_1}^2+m_{\tilde{t}_2}^2}\lesssim
\sqrt{3}$~\cite{Blinov:2013fta} (see also the earlier analysis to give
the bound $X_t /\sqrt{m_{\tilde{t}_1}^2+m_{\tilde{t}_2}^2}\lesssim
\sqrt{7}$~\cite{Kusenko:1996jn}.)\footnote{We have checked that in
  this region of $X_t$ the $\rho$ parameter is within the $2\sigma$
  bound of the observed value $\Delta\rho=(4.2\pm2.7)\times
  10^{-4}$~\cite{Barger:2012hr} based on
  Refs.\,\cite{Lim:1983re,Drees:1990dx,Heinemeyer:2004gx,Pierce:2016nwg}. It
  is also confirmed that Higgs-gluon-gluon coupling is within
  25\%~\cite{Khachatryan:2016vau} of the SM value referring to
  Refs.\,\cite{Djouadi:2005gi,Djouadi:2005gj}.}

It is seen that the deviation increases monotonically as $\sqrt{s}$
gets large for fixed $m_{L,R}$ and $X_t$. It is attributed to the
logarithmic term (first term of
Eq.\,\eqref{eq:stoploop_highenergylim_GpGp}), which originates in the
stop-sbottom loop and can be understood by RG running of the Higgs
quartic coupling.\footnote{The calculation is valid in a much higher
  $\sqrt{s}$ value, and a larger deviation is given in the energy
  range. However, it might be unnecessary information for a realistic
  (future) collider search; thus, we have omitted it.}  A smaller
$m_{L,R}$ gives a larger deviation. For example, $\Delta_{G^+G^+}=$16
(28)\%, 7 (15)\%, and 2 (6)\% for $\sqrt{s}=5$ (10) TeV for
$m_{L,R}=0.5$, 1, and 2 TeV with $X_t=0.5m_L$, respectively. This is
because the $\log(m^2_{\tilde{t}}/\sqrt{tu})$ term, which is dominant
in ${\cal A}_{G^+G^+}^{\tilde{t}\mathchar`-{\tilde{b}}}$ (see
Eq.\,\eqref{eq:stoploop_highenergylim_GpGp}), contributes
constructively in the total amplitude for $\sqrt{s}>m_{\tilde{t}}$.
It is true for the split mass spectrum (right panel).

When $X_t=X_t^{\rm fit}$, on the other hand, $\Delta_{G^+G^+}$ gets
smaller compared to the result with the same $m_{L,R}$ but
$X_t=0.5m_L$.  To understand the behavior, we plot $\Delta_{G^+G^+}$
as a function of $X_t$ for various $\sqrt{s}$ in
Fig.\,\ref{fig:Delta-Xt_GpGp} for $m_L=m_R=1$ TeV (left) and
$m_L=2m_R=1$ TeV (right). It is found that $\Delta_{G^+G^+}$ decreases
as $X_t$ increases for $X_t\lesssim X_t^{\rm fit}$, which can be
understood from Eqs.\,\eqref{eq:DeltaAnum} (or \eqref{eq:DeltaA}) and
\eqref{eq:stoploop_highenergylim_GpGp}. The second term on the
right-hand side of Eq.\,\eqref{eq:DeltaAnum} is positive and
destructively interferes with ${\cal
  A}_{G^+G^+}^{\tilde{t}\mathchar`-{\tilde{b}}}$ for $X_t\lesssim
X_t^{\rm fit} (\sim m_{\tilde{t}})$.

For larger $|\cos \theta|$ the deviation gets smaller since $Z$ and
$\gamma$ exchange terms which are proportional to $1/u$ or $1/t$
dominate the scattering amplitude. For example, when $\cos \theta =
0.5$, $\Delta_{G^+G^+}=$ 11\,(18)\%, 4.8\,(10)\%, and 2\,(4)\% for
$\sqrt{s}=5$\,(10) TeV $m_L=m_R=0.5$, 1, and 2 TeV with $X_t=0.5m_L$,
respectively.

$G^+G^-$ scattering has similar behavior except for a bump, which
corresponds to a resonance at $\sqrt{s}\simeq 2m_{\tilde{t}}$ in
Fig.\,\ref{fig:Delta-rs_GpGm}. This bump is due to discontinuity of
the first derivative of $B_0(q^2,m_1^2,m_2^2)$ with respect to $q^2$
at $q^2\simeq (m_1+m_2)^2$. For $\sqrt{s}\lesssim 2 m_{\tilde{t}}$,
the stop-sbottom loop (dominated by circle diagrams) is positive,
which constructively interferes with tree plus top-bottom loop
contributions. In the region $\sqrt{s}\gtrsim 2 m_{\tilde{t}}$, on the
other hand, $\Delta_{G^+G^-}$ monotonically decreases. This is because
the dominant term $\log(m^2_{\tilde{t}}/\sqrt{-st})$ in
Eq.\,\eqref{eq:stoploop_highenergylim_GpGm} is negative, which is
destructive in the total amplitude at the high energy range. For
example, $\Delta_{G^+G^-}=-29$ ($-42$)\%, $-11$ ($-25$)\%, and $6$
($-9$)\% for $\sqrt{s}=5$ (10) TeV and $\cos \theta=0.5$ for
$m_L=m_R=0.5$, 1, and 2 TeV with $X_t=0.5m_L$, respectively (left
panel). Qualitatively the same behavior is seen for the split mass
case (right panel). Regarding $X_t$ dependence, it is seen that
$|\Delta_{G^+G^-}|$ gets smaller for $X_t=X_t^{\rm fit}$ similarly to
the $G^+G^+$ case. Fig.\,\ref{fig:Delta-Xt_GpGm} clarifies the
behavior. It is found that $|\Delta_{G^+G^-}|$ decreases in the
$X_t\lesssim m_{\tilde{t}}$ region, which to attributed to the second
term on the right-hand side of Eq.\,\eqref{eq:DeltaAnum} as explained
in the $G^+G^+$ case.

Thus, in both the $G^+G^+$ and $G^+G^-$ scattering processes, it would be
difficult to observe the deviation from the SM in the parameter space
$|X_t|\sim m_{\tilde{t}}$, especially $X_t\simeq X_t^{\rm fit}$, since
$\Delta_{G^+G^\pm}$ is a few percent. In other words, scenario (a) is like
a ``blind spot'' for the TeV-scale stop search in the longitudinal $W$
boson scattering processes. In scenario (b), on the other hand, ${\cal
  O}$(1--10\%) deviation is expected for $\sqrt{s}=1$--$10$~TeV.

\section{concluding remarks}
\label{sec:conclusion}

In this paper we have studied high energy longitudinal $W$ boson
scattering with a light scalar top of which the mass is a few hundred
GeV to a few TeV. They affect the SM Higgs potential at quantum level,
and consequently the deviation from the standard model in longitudinal
gauge boson scattering is expected from the equivalence theorem.
Applying the equivalence theorem, we have computed charged
Nambu-Goldstone boson scattering processes and substituted them as
high energy $W^+_LW^{\pm}_L$ scattering processes.  In the study, we
consider two scenarios: (a) Higgs mass is explained in the MSSM
particle contents, and (b) other contributions besides the MSSM
particles make the observed Higgs mass.  It has been found that ${\cal
  O}$(1--10\%) deviation in the differential cross section is
predicted depending on stop mass and kinematics. As an example of
scenario (b), for $\sqrt{s}=5$ (10) TeV and $\cos \theta =0$, the
deviation in the $W^+_LW^+_L$ process is 16 (28)\% and 7 (15)\% when
both left- and right-handed stop masses ($m_L$ and $m_R$) are 0.5 and
1 TeV with the mixing parameter $X_t=0.5m_L$, respectively. Similarly,
in $W^+_LW^-_L$, it is $-29$ ($-42$)\%, and $-11$ ($-25$)\% but for
$\cos \theta =0.5$.  For scenario (a), on the contrary, it has been discovered
that the deviation gets smaller, {\it e.g.}, $2$ ($4$)\% and $4$
($-4$)\% for $m_L=m_R=1$ TeV with the appropriate $X_t$ for
$\sqrt{s}=5$ (10) in $W^+_LW^+_L$ and $W^+_LW^-_L$, respectively. The
same behavior is seen for the $m_L\neq m_R$ case. Thus in such a case it
would be challenging to see the existence of stop in $W_LW_L$
scattering.

High energy longitudinal gauge boson scattering has started to be
measured at the LHC~\cite{Aad:2014zda,Khachatryan:2014sta}. However,
the observation of ${\cal O}(10\%)$ deviation would be difficult even
in Run 2 at the LHC. This is because the number of events which has
over a few TeV invariant mass of $W$ boson system is suppressed due to
gauge cancellation~\cite{Accomando:2006mc}. (We have checked this by
using the MadGraph package~\cite{Alwall:2014hca}.\footnote{We thank
  Yasuhiro Shimizu for pointing this out and information useful for
  performing MadGraph5.}) Thus at least an upgraded program, such as
the High Luminosity LHC, would be necessary. Or the Future Circular
Collider, which is planed to operate at 100 TeV center-of-mass energy,
would be more promising for the study of the gauge boson
scattering. In such a high energy experiment, the observation of stop
or sbottom pair production might be more direct and easier way to
observe a clue of the supersymmetry. As mentioned in the Introduction,
however, there are model dependence in the data analysis, {\it e.g.},
details of the decay modes, or violation of R-parity.  High energy
longitudinal gauge boson scattering would be complementary to the
direct searches.  We have provided the theoretical ingredients for the
numerical study and discuss feasibility for the discovery of scalar
tops in the longitudinal gauge boson scattering. The next step will be
to perform full simulation for hadron or lepton collider experiments
with various energies, for which
Refs.\,\cite{Denner:1996ug,Denner:1997kq,Biedermann:2016yds,Borel:2012by,Alboteanu:2008my,Accomando:2006hq,Bernreuther:2015llj,Fleper:2016frz,Bishara:2016kjn,Doroba:2012pd,Kilian:2014zja,Kilian:2015opv}
are useful.  We leave it to future work.

\vspace{0.3cm}

\noindent
{\it Acknowledgements}

\noindent    
We are grateful to Kazuhiro Endo and Yukinari Sumino for valuable
discussions. This work was supported by MEXT KAKENHI Grant Number
17H05402 and JSPS KAKENHI Grant Number 17K14278. (K.I.)


\appendix

\section{Analytic check}
\label{app:analyticcheck}

We will check that the $G^+G^+$ and $G^+G^-$ scattering amplitudes
reduce to those in the SM in the low-energy limit by using the
analytic Higgs mass formula for the $\tilde{m}\sim m_{\tilde{t}}$
case.

First of all we need to use the Lagrangian after the following
replacement:
\begin{eqnarray}
  \lambda_{\rm H}^{\rm SM'} \to (1/8)g_Z^2 \cos^22\beta \,,
  \,\,\,\,
  y_t^{\rm SM'} \to \lambda_t\sin\beta\,.
\end{eqnarray}
Consequently, the matching conditions are
\begin{align}
 &\lambda_{\rm H}^{\rm SM}(\mu_{\tilde{t}})=
   \frac{1}{8}g_Z^2 \cos^22\beta(\mu_{\tilde{t}})
  \nonumber \\
  &\qquad + \frac{N_C (y_t^{{\rm SM'}}(\mu_{\tilde{t}}))^4}{(4\pi)^2}
  \Bigl[-\log \Bigl(\frac{\mu^2_{\tilde{t}}}{m_{\tilde{t}}^2}\Bigr)
    +\frac{X_t^2}{m_{\tilde{t}}^2}
    \Bigl(1-\frac{X_t^2}{12m_{\tilde{t}}^2}\Bigr)\Bigr]\,, 
  \label{eq:matching_lambdaH_susy} \\
  &y_t^{\rm SM}(\mu_{\tilde{t}})=\lambda_t\sin\beta (\mu_{\tilde{t}})\,.
  \label{eq:matching_yt_susy}
\end{align}
  
In the MSSM where $\tilde{m}\sim m_{\tilde{t}}$, the SM Higgs mass is
given by analytically using the effective
potential~\cite{Haber:1990aw,Ellis:1990nz,
  Ellis:1991zd,Okada:1990vk}\footnote{For review, see
  Ref.\,\cite{Drees_textbook}. For diagrammatic calculation, see, {\it
    e.g.}, Ref.\,\cite{Brignole:1992uf}. It is shown that the
  diagrammatic calculation well agrees with the result in the effective
  potential approach.}
\begin{align}
 &(m_h^{{\rm MSSM}})^2\simeq m_Z^2\cos^22\beta
  \nonumber \\
  &+ \frac{2N_c (\lambda_t\sin\beta)^4 v^2}{(4\pi)^2}
  \Bigl[\log \Bigl(\frac{m_{\tilde{t}}^2}{m_t^2}\Bigr)
    +\frac{X_t^2}{m_{\tilde{t}}^2}
    \Bigl(1-\frac{X_t^2}{12m_{\tilde{t}}^2}\Bigr)\Bigr]\,.
\label{eq:mhMSSM}
\end{align}
Using this expression, it is found that the $X_t$ dependence on
$m_h^{{\rm MSSM}}$ and the one obtained by using the RG equation in
the text agree within around 1~GeV when we take $\mu=m_{\tilde{t}}$
for top Yukawa coupling. Hereafter, we take $\mu=m_{\tilde{t}}$.

$G^+G^+$ and $G^+G^-$ scattering amplitudes are easily obtained by
using the previous result along with the above replacement and
matching conditions.  Now let us see the low-energy limit,
$s,|t|,|u|\ll m_{\tilde{t}}^2$ (but $s,|t|,|u|\gg m_Z^2$).  ${\cal
  A}_{G^+G^\pm}^{\tilde{t}\mathchar`-{\tilde{b}}}$ corresponding to
Eq.\,\eqref{eq:A^stop_lowenergylim} is the same expression.  Then,
combining with
\begin{align}
  &{\cal A}_{G^+G^+}^{t\mathchar`-b} \longrightarrow
  -\frac{4N_c (y_t^{\rm SM})^4}{(4\pi)^2}
  \Bigl[
    \log\Bigl(\frac{\mu^2}{\sqrt{tu}}\Bigr)+2
    \Bigr]\Bigr|_{\mu=m_{\tilde{t}}}\,,
  \\
  &{\cal A}_{G^+G^-}^{t\mathchar`-b} \longrightarrow
  -\frac{4N_c (y_t^{\rm SM})^4}{(4\pi)^2}
  \Bigl[
    \log\Bigl(\frac{\mu^2}{\sqrt{st}}\Bigr)+2
    \Bigr]\Bigr|_{\mu=m_{\tilde{t}}}\,,
\end{align}
and Eq.\,\eqref{eq:mhMSSM}, we obtain
\begin{align}
  {\cal A}_{G^+G^+} \longrightarrow &
  -\frac{2(m_h^{{\rm MSSM}})^2}{v^2}
  -\frac{g_Z^2}{2}\Bigl[\frac{t}{u}+\frac{u}{t}+1\Bigr]
  \nonumber \\ &-
  \frac{2N_c (y_t^{\rm SM})^4}{(4\pi)^2}
  \Bigl[
    \log\Bigl(\frac{m_t^4}{tu}\Bigr)+4
    \Bigr]\,,
  \\
  {\cal A}_{G^+G^-} \longrightarrow &
  -\frac{2(m_h^{{\rm MSSM}})^2}{v^2}
  -\frac{g_Z^2}{2}\Bigl[\frac{s}{t}+\frac{t}{s}+1\Bigr]
  \nonumber \\ &-
  \frac{2N_c (y_t^{\rm SM})^4}{(4\pi)^2}
  \Bigl[
    \log\Bigl(\frac{m_t^4}{st}\Bigr)+4
    \Bigr]\,. 
\end{align}
This is exactly the amplitude including top-bottom loop in the SM for
$m_{h}^{\rm MSSM}=m_h$~\cite{Endo:2016koi}.


\end{document}